%Paper: cond-mat/9310053
%From: mvekic@isis.ps.uci.edu (Marco Vekic)
%Date: Fri, 22 Oct 93 14:34:08 PDT

%% FOLLOWING LINE CANNOT BE BROKEN BEFORE 80 CHAR
%%%%%%%%%%%%%%%%%%%%%%%%%%%%%%%%%%%%%%%%%%%%%%%%%%%%%%%%%%%%%%%%%%%%%%%%%%%%%%%%
% Create separate files for cjnl.tex, reforder.tex, eqnorder.tex, smooth.tex.
% Use plain TeX.
%% FOLLOWING LINE CANNOT BE BROKEN BEFORE 80 CHAR
%%%%%%%%%%%%%%%%%%%%%%%%%%%%%%%%%%%%%%%%%%%%%%%%%%%%%%%%%%%%%%%%%%%%%%%%%%%%%%%%
%%                      CJNL.TEX
%% FOLLOWING LINE CANNOT BE BROKEN BEFORE 80 CHAR
%%%%%%%%%%%%%%%%%%%%%%%%%%%%%%%%%%%%%%%%%%%%%%%%%%%%%%%%%%%%%%%%%%%%%%%%%%%%%%%%
%%
%%                This is CJNL.TEX Version 0.3 as of 6/12/85.
%%
%%	This is a set of TeX 82 macros designed to produce scientific
%%	papers with a minimum of fuss and using as much of plain.tex as
%%	possible.  The user need only know what is in the TeXbook, and
%%	the macros under ``user definitions'' below.  Also, the user
%%	definitions are intended to be as simple as possible, so that
%%	the user may change them as desired.

%%
%%  Font definitions suitable for the IMAGEN (Written by Tony Kennedy)
%%

%  Define a whole menagerie of pseudo-12pt fonts

\font\twelverm=cmr10 scaled 1200    \font\twelvei=cmmi10 scaled 1200
\font\twelvesy=cmsy10 scaled 1200   \font\twelveex=cmex10 scaled 1200
\font\twelvebf=cmbx10 scaled 1200   \font\twelvesl=cmsl10 scaled 1200
\font\twelvett=cmtt10 scaled 1200   \font\twelveit=cmti10 scaled 1200

\skewchar\twelvei='177   \skewchar\twelvesy='60

%  Define \...point macros to change fonts and spacings consistently

\def\twelvepoint{\normalbaselineskip=12.4pt
  \abovedisplayskip 12.4pt plus 3pt minus 9pt
  \belowdisplayskip 12.4pt plus 3pt minus 9pt
  \abovedisplayshortskip 0pt plus 3pt
  \belowdisplayshortskip 7.2pt plus 3pt minus 4pt
  \smallskipamount=3.6pt plus1.2pt minus1.2pt
  \medskipamount=7.2pt plus2.4pt minus2.4pt
  \bigskipamount=14.4pt plus4.8pt minus4.8pt
  \def\rm{\fam0\twelverm}          \def\it{\fam\itfam\twelveit}%
  \def\sl{\fam\slfam\twelvesl}     \def\bf{\fam\bffam\twelvebf}%
  \def\mit{\fam 1}                 \def\cal{\fam 2}%
  \def\tt{\twelvett}
  \def\nullspace{\nulldelimiterspace=0pt \mathsurround=0pt }
  \def\big##1{{\hbox{$\left##1\vbox to 10.2pt{}\right.\nullspace$}}}
  \def\Big##1{{\hbox{$\left##1\vbox to 13.8pt{}\right.\nullspace$}}}
  \def\bigg##1{{\hbox{$\left##1\vbox to 17.4pt{}\right.\nullspace$}}}
  \def\Bigg##1{{\hbox{$\left##1\vbox to 21.0pt{}\right.\nullspace$}}}
  \textfont0=\twelverm   \scriptfont0=\tenrm   \scriptscriptfont0=\sevenrm
  \textfont1=\twelvei    \scriptfont1=\teni    \scriptscriptfont1=\seveni
  \textfont2=\twelvesy   \scriptfont2=\tensy   \scriptscriptfont2=\sevensy
  \textfont3=\twelveex   \scriptfont3=\twelveex  \scriptscriptfont3=\twelveex
  \textfont\itfam=\twelveit
  \textfont\slfam=\twelvesl
  \textfont\bffam=\twelvebf \scriptfont\bffam=\tenbf
  \scriptscriptfont\bffam=\sevenbf
  \normalbaselines\rm}

%	tenpoint

%%
%%	Various internal macros
%%

\def\beginlinemode{\endmode
  \begingroup\parskip=0pt \obeylines\def\\{\par}\def\endmode{\par\endgroup}}
\def\beginparmode{\endmode
  \begingroup \def\endmode{\par\endgroup}}
\let\endmode=\par
{\obeylines\gdef\
{}}
\def\singlespace{\baselineskip=\normalbaselineskip}

\def\oneandahalfspace{\baselineskip=\normalbaselineskip
  \multiply\baselineskip by 3 \divide\baselineskip by 2}
\def\doublespace{\baselineskip=\normalbaselineskip \multiply\baselineskip by 2}

\newcount\firstpageno
\firstpageno=2
%% FOLLOWING LINE CANNOT BE BROKEN BEFORE 80 CHAR
\footline={\ifnum\pageno<\firstpageno{\hfil}\else{\hfil\twelverm\folio\hfil}\fi}
\let\rawfootnote=\footnote		% We must set the footnote style
\def\footnote#1#2{{\rm\singlespace\parindent=0pt\rawfootnote{#1}{#2}}}
\def\raggedcenter{\leftskip=4em plus 12em \rightskip=\leftskip
  \parindent=0pt \parfillskip=0pt \spaceskip=.3333em \xspaceskip=.5em
  \pretolerance=9999 \tolerance=9999
  \hyphenpenalty=9999 \exhyphenpenalty=9999 }
\def\dateline{\rightline{\ifcase\month\or
  January\or February\or March\or April\or May\or June\or
  July\or August\or September\or October\or November\or December\fi
  \space\number\year}}
\def\received{\vskip 3pt plus 0.2fill
 \centerline{\sl (Received\space\ifcase\month\or
  January\or February\or March\or April\or May\or June\or
  July\or August\or September\or October\or November\or December\fi
  \qquad, \number\year)}}

%%
%%	Page layout, margins, font and spacing (feel free to change)
%%

\hsize=6.5truein
\hoffset=0truein
\vsize=8.9truein
\voffset=0truein
\parskip=\medskipamount
\twelvepoint		% selects twelvepoint fonts (cf. \tenpoint)
\doublespace		% selects double spacing for main part of paper (cf.
			%	\singlespace, \oneandahalfspace)
\overfullrule=0pt	% delete the nasty little black boxes for overfull box

%%
%%	The user definitions for major parts of a paper (feel free to change)
%%

	% Preprint number at upper right of title page

\def\title			%  Title on title page
  {\null\vskip 3pt plus 0.2fill
   \beginlinemode \doublespace \raggedcenter \bf}

\def\author			%  Author(s) name(s)  on title page
  {\vskip 3pt plus 0.2fill \beginlinemode
   \singlespace \raggedcenter}

\def\affil			% Affiliations (can intermix with \author)
  {\vskip 3pt plus 0.1fill \beginlinemode
   \oneandahalfspace \raggedcenter \sl}

\def\abstract			% Begin abstract
  {\vskip 3pt plus 0.3fill \beginparmode
   \doublespace \narrower ABSTRACT: }

\def\endtitlepage		% End title page, begin body of paper
  {\endpage			% 	This subsumes \body
   \body}

\def\body			% Begin text body;  can be used to end
  {\beginparmode}		% \title, \author, \affil, \abstract,
				% \reference, or \figurecaption modes

\def\head#1{			% Head;  NOTE enclose the text in {}
  \filbreak\vskip 0.5truein	%  e.g., \head{I. Introduction}
  {\immediate\write16{#1}
   \raggedcenter \uppercase{#1}\par}
   \nobreak\vskip 0.25truein\nobreak}

\def\refto#1{$^{#1}$}		% For references in text as superscript

\def\references			% Begin references -- basic format is Phys Rev
  {\endpage
   \head{References}		% I.e., volume, page, year (space after commas).
   \beginparmode
   \frenchspacing \parindent=0pt \leftskip=1truecm
   \parskip=8pt plus 3pt \everypar{\hangindent=\parindent}}

\gdef\refis#1{\indent\hbox to 0pt{\hss#1.~}}	% Ref list numbers.

\gdef\journal#1, #2, #3, 1#4#5#6{		% Journal reference.  Comma sets
    {\sl #1~}{\bf #2}, #3, (1#4#5#6)}		% off: name, vol, page, year

\gdef\journ2 #1, #2, #3, 1#4#5#6{		% Journal reference.  Comma sets
    {\sl #1~}{\bf #2}: #3, (1#4#5#6)}		% off: name, vol, page, year
                                     		% Colon inserted after volume #

\def\refstylenp{		% Nucl Phys(or Phys Lett) ref style: V, Y, P
  \gdef\refto##1{ [##1]}				% Reference in text []
  \gdef\refis##1{\indent\hbox to 0pt{\hss##1)~}}	% Ref list numbers)
  \gdef\journal##1, ##2, ##3, ##4 {			% Journal reference
     {\sl ##1~}{\bf ##2~}(##3) ##4 }}

\def\refstyleprnp{		% Input like pr, output like np!!
  \gdef\refto##1{ [##1]}				% Reference in text []
  \gdef\refis##1{\indent\hbox to 0pt{\hss##1)~}}	% Ref list numbers)
  \gdef\journal##1, ##2, ##3, 1##4##5##6{		% Journal reference
    {\sl ##1~}{\bf ##2~}(1##4##5##6) ##3}}

\def\prl{\journal Phys. Rev. Lett., }

\def\endreferences{\body}

\def\figurecaptions		% Begin figure captions
  {\endpage
   \beginparmode
   \head{Figure Captions}
}

\def\endfigurecaptions{\body}

\def\tables             % Begin tables
  {\endpage
   \beginparmode
   \head{Tables}
}

\def\endpage			%  Eject a page
  {\vfill\eject}

\def\endpaper			%  Ways to say goodbye
  {\endmode\vfill\supereject}

%%
%%	Various little user definitions
%%

\def\ref#1{Ref. #1}			% 	for inline references
			% 	ditto

		% For citation of equation numbers
	%	ditto
			%	ditto
			%	ditto
			%	ditto
			%	ditto
\def\frac#1#2{{\textstyle #1 \over \textstyle #2}}

\def\sla{\raise.15ex\hbox{$/$}\kern-.57em}
\def\leaderfill{\leaders\hbox to 1em{\hss.\hss}\hfill}
\def\twiddle{\lower.9ex\rlap{$\kern-.1em\scriptstyle\sim$}}
\def\bigtwiddle{\lower1.ex\rlap{$\sim$}}
\def\gtwid{\mathrel{\raise.3ex\hbox{$>$\kern-.75em\lower1ex\hbox{$\sim$}}}}
\def\ltwid{\mathrel{\raise.3ex\hbox{$<$\kern-.75em\lower1ex\hbox{$\sim$}}}}
\def\square{\kern1pt\vbox{\hrule height 1.2pt\hbox{\vrule width 1.2pt\hskip 3pt
   \vbox{\vskip 6pt}\hskip 3pt\vrule width 0.6pt}\hrule height 0.6pt}\kern1pt}

\catcode`@=11
\newcount\tagnumber\tagnumber=0

\immediate\newwrite\eqnfile
\newif\if@qnfile\@qnfilefalse
\def\write@qn#1{}
\def\writenew@qn#1{}
\def\w@rnwrite#1{\write@qn{#1}\message{#1}}
\def\@rrwrite#1{\write@qn{#1}\errmessage{#1}}

\def\taghead#1{\gdef\t@ghead{#1}\global\tagnumber=0}
\def\t@ghead{}

\expandafter\def\csname @qnnum-3\endcsname
  {{\t@ghead\advance\tagnumber by -3\relax\number\tagnumber}}
\expandafter\def\csname @qnnum-2\endcsname
  {{\t@ghead\advance\tagnumber by -2\relax\number\tagnumber}}
\expandafter\def\csname @qnnum-1\endcsname
  {{\t@ghead\advance\tagnumber by -1\relax\number\tagnumber}}
\expandafter\def\csname @qnnum0\endcsname
  {\t@ghead\number\tagnumber}
\expandafter\def\csname @qnnum+1\endcsname
  {{\t@ghead\advance\tagnumber by 1\relax\number\tagnumber}}
\expandafter\def\csname @qnnum+2\endcsname
  {{\t@ghead\advance\tagnumber by 2\relax\number\tagnumber}}
\expandafter\def\csname @qnnum+3\endcsname
  {{\t@ghead\advance\tagnumber by 3\relax\number\tagnumber}}

\def\equationfile{%
  \@qnfiletrue\immediate\openout\eqnfile=\jobname.eqn%
  \def\write@qn##1{\if@qnfile\immediate\write\eqnfile{##1}\fi}
  \def\writenew@qn##1{\if@qnfile\immediate\write\eqnfile
    {\noexpand\tag{##1} = (\t@ghead\number\tagnumber)}\fi}
}

\def\callall#1{\xdef#1##1{#1{\noexpand\call{##1}}}}
\def\call#1{\each@rg\callr@nge{#1}}

\def\each@rg#1#2{{\let\thecsname=#1\expandafter\first@rg#2,\end,}}
\def\first@rg#1,{\thecsname{#1}\apply@rg}
\def\apply@rg#1,{\ifx\end#1\let\next=\relax%
\else,\thecsname{#1}\let\next=\apply@rg\fi\next}

\def\callr@nge#1{\calldor@nge#1-\end-}
\def\callr@ngeat#1\end-{#1}
\def\calldor@nge#1-#2-{\ifx\end#2\@qneatspace#1 %
  \else\calll@@p{#1}{#2}\callr@ngeat\fi}
\def\calll@@p#1#2{\ifnum#1>#2{\@rrwrite{Equation range #1-#2\space is bad.}
\errhelp{If you call a series of equations by the notation M-N, then M and
N must be integers, and N must be greater than or equal to M.}}\else%
 {\count0=#1\count1=#2\advance\count1
by1\relax\expandafter\@qncall\the\count0,%
  \loop\advance\count0 by1\relax%
    \ifnum\count0<\count1,\expandafter\@qncall\the\count0,%
  \repeat}\fi}

\def\@qneatspace#1#2 {\@qncall#1#2,}
\def\@qncall#1,{\ifunc@lled{#1}{\def\next{#1}\ifx\next\empty\else
  \w@rnwrite{Equation number \noexpand\(>>#1<<) has not been defined yet.}
  >>#1<<\fi}\else\csname @qnnum#1\endcsname\fi}

\let\eqnono=\eqno
\def\eqno(#1){\tag#1}
\def\tag#1$${\eqnono(\displayt@g#1 )$$}

\def\aligntag#1\endaligntag
  $${\gdef\tag##1\\{&(##1 )\cr}\eqalignno{#1\\}$$
  \gdef\tag##1$${\eqnono(\displayt@g##1 )$$}}

\def\eqalignno#1{\displ@y \tabskip\centering
  \halign to\displaywidth{\hfil$\displaystyle{##}$\tabskip\z@skip
    &$\displaystyle{{}##}$\hfil\tabskip\centering
    &\llap{$\displayt@gpar##$}\tabskip\z@skip\crcr
    #1\crcr}}

\def\displayt@gpar(#1){(\displayt@g#1 )}

\def\displayt@g#1 {\rm\ifunc@lled{#1}\global\advance\tagnumber by1
        {\def\next{#1}\ifx\next\empty\else\expandafter
        \xdef\csname @qnnum#1\endcsname{\t@ghead\number\tagnumber}\fi}%
  \writenew@qn{#1}\t@ghead\number\tagnumber\else
        {\edef\next{\t@ghead\number\tagnumber}%
        \expandafter\ifx\csname @qnnum#1\endcsname\next\else
        \w@rnwrite{Equation \noexpand\tag{#1} is a duplicate number.}\fi}%
  \csname @qnnum#1\endcsname\fi}

\def\ifunc@lled#1{\expandafter\ifx\csname @qnnum#1\endcsname\relax}

\let\@qnend=\end\gdef\end{\if@qnfile
\immediate\write16{Equation numbers written on []\jobname.EQN.}\fi\@qnend}

\catcode`@=12

\catcode`@=11
\newcount\r@fcount \r@fcount=0
\newcount\r@fcurr
\immediate\newwrite\reffile
\newif\ifr@ffile\r@ffilefalse
\def\w@rnwrite#1{\ifr@ffile\immediate\write\reffile{#1}\fi\message{#1}}

\def\writer@f#1>>{}
\def\referencefile{%			  Stuff to write .REF file
  \r@ffiletrue\immediate\openout\reffile=\jobname.ref%
  \def\writer@f##1>>{\ifr@ffile\immediate\write\reffile%
    {\noexpand\refis{##1} = \csname r@fnum##1\endcsname = %
     \expandafter\expandafter\expandafter\strip@t\expandafter%
     \meaning\csname r@ftext\csname r@fnum##1\endcsname\endcsname}\fi}%
  \def\strip@t##1>>{}}

\def\citeall#1{\xdef#1##1{#1{\noexpand\cite{##1}}}}
\def\cite#1{\each@rg\citer@nge{#1}}	% Variable No. of args, separated by ","

\def\each@rg#1#2{{\let\thecsname=#1\expandafter\first@rg#2,\end,}}
\def\first@rg#1,{\thecsname{#1}\apply@rg}	% each@ag is a general purpose
\def\apply@rg#1,{\ifx\end#1\let\next=\relax%	  variable no. of arg. macro.
\else,\thecsname{#1}\let\next=\apply@rg\fi\next}% args separated by commas

\def\citer@nge#1{\citedor@nge#1-\end-}	% Check for M-N range (M and N numbers)
\def\citer@ngeat#1\end-{#1}
\def\citedor@nge#1-#2-{\ifx\end#2\r@featspace#1 % Single argument
  \else\citel@@p{#1}{#2}\citer@ngeat\fi}	% M-N range of arguments
\def\citel@@p#1#2{\ifnum#1>#2{\errmessage{Reference range #1-#2\space is bad.}
    \errhelp{If you cite a series of references by the notation M-N, then M and
    N must be integers, and N must be greater than or equal to M.}}\else%
 {\count0=#1\count1=#2\advance\count1
by1\relax\expandafter\r@fcite\the\count0,%
  \loop\advance\count0 by1\relax%	  Loop from M to N
    \ifnum\count0<\count1,\expandafter\r@fcite\the\count0,%
  \repeat}\fi}

\def\r@featspace#1#2 {\r@fcite#1#2,}	% Eat spaces at beginning or end of arg
\def\r@fcite#1,{\ifuncit@d{#1}		% Cite individual reference
    \expandafter\gdef\csname r@ftext\number\r@fcount\endcsname%
    {\message{Reference #1 to be supplied.}\writer@f#1>>#1 to be supplied.\par
     }\fi%
  \csname r@fnum#1\endcsname}

\def\ifuncit@d#1{\expandafter\ifx\csname r@fnum#1\endcsname\relax%
\global\advance\r@fcount by1%
\expandafter\xdef\csname r@fnum#1\endcsname{\number\r@fcount}}

\let\r@fis=\refis			% Save old \refis, redefine
\def\refis#1#2#3\par{\ifuncit@d{#1}%      Use two params #2 #3 to strip blank
    \w@rnwrite{Reference #1=\number\r@fcount\space is not cited up to now.}\fi%
  \expandafter\gdef\csname r@ftext\csname r@fnum#1\endcsname\endcsname%
  {\writer@f#1>>#2#3\par}}

\def\r@ferr{\endreferences\errmessage{I was expecting to see
\noexpand\endreferences before now;  I have inserted it here.}}
\let\r@ferences=\references
\def\references{\r@ferences\def\endmode{\r@ferr\par\endgroup}}

\let\endr@ferences=\endreferences
\def\endreferences{\r@fcurr=0%		  Save old \endreferences, redefine
  {\loop\ifnum\r@fcurr<\r@fcount%	  Loop over refnum and produce text
    \advance\r@fcurr by 1\relax\expandafter\r@fis\expandafter{\number\r@fcurr}%
    \csname r@ftext\number\r@fcurr\endcsname%
  \repeat}\gdef\r@ferr{}\endr@ferences}

% Save old \endpaper, redefine it to write parting message.

\let\r@fend=\endpaper\gdef\endpaper{\ifr@ffile
\immediate\write16{Cross References written on []\jobname.REF.}\fi\r@fend}

\catcode`@=12

\citeall\refto		% These macros will generate citations
\citeall\ref		%
\citeall
%
%%%%%%%%%%%%%%%%%%%%%%%%%%%%%%%%%%%%%%%%%%%%%%%%%%%%%%%%%%%%%%%%%%%%%%%%%%%%%%%
%%                            SMOOTH.TEX
%%%%%%%%%%%%%%%%%%%%%%%%%%%%%%%%%%%%%%%%%%%%%%%%%%%%%%%%%%%%%%%%%%%%%%%%%%%%%%%
\input cjnl
\input reforder
\input eqnorder
%\nopagenumbers
\def\(#1){(\call{#1})}

\def\Vekic{Veki\'c}
\def\bra{\langle}
\def\ket{\rangle}

\rightline{October, 1993}
\title Smooth Boundary Conditions For Quantum Lattice Systems

\author M.~\Vekic~and~S.R.~White
\bigskip
\centerline{\sl Department of Physics}
\centerline{\sl University of California}
\centerline{\sl Irvine, CA  92717}

\abstract
We introduce a new type of boundary conditions,
{\it smooth boundary conditions}, for numerical studies of
quantum lattice systems. In a number of circumstances, these boundary
conditions have substantially smaller finite-size effects than periodic
or open boundary conditions.
They can be applied to nearly any short-ranged Hamiltonian system
in any dimensionality and within almost any type of
numerical approach.

\noindent PACS Numbers: 02.70.+d, 05.30.Fk, 75.10.Jm
\endtitlepage

\vfill\eject
\centerline {\bf I. Introduction}

In most numerical calculations for quantum systems, periodic boundary
conditions (PBC's) are the accepted standard. There are a number of
situations, however, where PBC's are inadequate. In systems with some form
of incommensurate order, for example, very large system sizes are needed
to approximate the incommensurate behavior of the infinite system,
and in mean-field methods, where a
number of iterations are required to achieve convergence, the system
can get stuck in a commensurate state far from the desired incommensurate
order. Another example, which forms the primary motivation for this work,
stems from the density-matrix renormalization group (RG)
method.\refto{RGWhite} This new
real-space numerical method has proven to be extremely accurate for
Heisenberg spin chains,\refto{RGWhite1}
but for greatest accuracy requires that the
chain {\it not} form a closed loop, as in PBC's. This poses no great
inconvenience for the $S=1$ chain, where there is a finite correlation
length, but is quite inconvenient for half-integer spin chains
(and most 1D fermion systems), where
boundary effects decay as a power law.

Recently, new types of boundary conditions, such as
self-determined boundary conditions
\refto{selfdetbc} and
nebula boundary conditions,\refto{nebulabc}
have been studied in conjunction with quantum Monte Carlo simulations,
but cannot be generalized
in an easy manner
to any arbitrary system or to other types of numerical techniques.
In this paper we introduce a new type of boundary conditions, {\it smooth
boundary conditions} (SBC's), which in the circumstances listed above,
perform better than PBC's and open boundary conditions (OBC's).
The main idea of these new boundary conditions is
to smoothly ``turn off" (set to zero) the parameters of the Hamiltonian
near the edges of the system. Surprisingly, in many cases where PBC's
or OBC's perform
very well, SBC's perform better. They can be applied to numerical calculations
for nearly any system with local interactions in any number of dimensions.

After introducing the ideas of SBC's, we will illustrate their use in several
systems. The ideas behind SBC's are closely related to the summation
of infinite series and the Borel transform, and we will motivate their
development by first discussing accelerated convergence of numerical series.

Let $s_n=\sum_{m=0}^n a_m$ be a slowly converging alternating series, with
$s =\lim_{n\rightarrow\infty} s_n$. For example, we can consider the series
$$
a_m = (-1)^m / \ln[\ln(m+3)] .
\eqno(loglog)
$$
The summation of such a series can be viewed as a termination problem;
if we stop with an odd number of terms, we get a positive result, while
stopping with an even number gives a negative result. We would like to find
some
way of terminating the series in a way that does not bias between odd and even
number of terms. We can do this by constructing a smoothing function,
$c_m$, and taking
$$
s \approx \sum_{m=0}^{M} a_m c_m.
\eqno(ssum)
$$
The smoothing function is conveniently described as a continuous function
$y(x)$, $0 \le x \le 1$, with $y(0)=1$ and $y(1)=0$,
sampled at a discrete set of $M$ points,
$$
c_m = y(m/M),
\eqno(cmdef)
$$
with $0<m<M$.
An effective choice for $y(x)$ is
$$
y(x) = {1 \over 2} \left[ 1 - \tanh {x-1/2 \over x(1-x)} \right] .
\eqno(ydef)
$$
This approach is remarkably successful at summing a wide variety of
common, slowly-converging, alternating series, such as those for
$\pi$, $\ln 2$, etc., attaining results accurate
to 10 or 12 digits with $~100$ terms. Convergence is roughly
exponential with $M$.  For the series in Eq.(\call{loglog}),
we obtain the result $s=8.749551241(2)$ with $M=100$.
The same $c_m$'s are used
for each series, and the total numerical work is extremely small.
Note that all derivatives of $y$ are zero at $0$ and $1$;
%$y(0)=1$, $y'(0)=y''(0)=\ldots=0$, $y(1)=y'(1)=y''(1)=\ldots=0$;
in fact, the function has essential singularities at 0 and 1.
These properties are crucial for effective termination of a series; for
example, if the function $y(x) = (1-x^2)^2$ is used, for which $y'' \ne 0$
at 0 and 1, convergence is only quadratic in $1/M$.

This procedure is closely related to the Borel transform,\refto{Hardy}
which is usually applied to {\it divergent} series. The Borel transform
of the series $s$ is defined as
$$
a(x) = \sum_{m=0}^\infty {a_m \over m!} x^m.
\eqno(borel)
$$
{}From the definition of $a(x)$ it follows trivially that
$$
s=\int_0^\infty dx e^{-x} a(x).
\eqno(borelint)
$$
The standard use of the Borel transform is to calculate $a(x)$ and then
perform the integration; however, here we will not calculate $a(x)$.
We will only assume that $a(x)e^{-x}$ is negligible for $x$ greater
than a cutoff $M'$. We take $M'$ as the upper limit of the integral
in Eq.(\call{borelint}), then replace $a(x)$ by its definition
Eq.(\call{borel}), and exchange the sum and integral.
We obtain
$$
s \approx \sum_{m=0}^{\infty} a_m c_m(M'),
\eqno(bsum)
$$
where
$$
c_m(M') = e^{-M'} \sum_{n=m+1}^\infty {M'^n \over n!}.
\eqno(borelc)
$$
For $m > M\approx 2 M'$, $c_m(M')$  is completely neglible,
and the sum in Eq.(\call{bsum}) can be terminated, yielding Eq.(\call{ssum}) .
In Fig.~1 we show both $c_m$ as defined in Eq.(\call{borelc}) with $M'=20$, and
$y(m/M)$ as defined in Eq.(\call{ydef}) with $M=40$.
The Borel approach and the approach using Eq.(\call{ydef}) are roughly
equally effective at summing common series. However, the Borel form
is slightly less convenient, since one must choose both $M$ and $M'$.

This approach to numerical series is largely pedagogical; there are
probably even more efficient ways to sum such series.
To apply these ideas to reduce finite-size effects in a general Hamiltonian
system, we consider first a trivial example,
a one-dimensional tight-binding chain. We consider an
$L$-site lattice with hopping matrix element
$t_i$, centered at $E=0$, with Fermi level $\epsilon_{\rm F}$, and Hamiltonian
matrix
$$
H_{ij} = -t_i \delta_{j,i+1} - t_j \delta_{i,j+1}.
\eqno(HoppingHam)
$$
Ordinarily $t_i$ (which gives the hopping between sites $i$ and $i+1$)
is a constant $t$. To apply SBC's, we
set
$$
t_i/t=
\cases{
c_{M-i}, & $1\le i\le M$;      \cr
1,         & $M < i\le L-M$; \cr
c_{i-L+M}, & $L-M < i < L$.\cr} \eqno(fsmooth)
$$
Here $c_i$ is the smoothing function defined by Eq.(\call{ydef}) or
Eq.(\call{borelc}).

We also need to adjust the diagonal elements of $H$. A general rule for
applying SBC's is that in the limit that the width of the smoothing region
$M\rightarrow\infty$, the local properties of the system should be constant
with $i$. In this case the Fermi level, $\epsilon_{\rm F}$,
is constant across the
system, so that, as we vary the local bandwidth, we must shift the band center
so that $\epsilon_{\rm F}$ strikes the band in the same {\it relative}
position.
Thus,
Eq.(\call{HoppingHam}) becomes
$$
H_{i,j} = -t_i \delta_{j,i+1} - t_j \delta_{i,j+1} +\delta_{i,j}\epsilon_{\rm
F}
\left(1-{{t_{i-1}+t_i}\over {2t}}\right).
\eqno(SoftHoppingHam)
$$
Note that Eq.(\call{SoftHoppingHam}) explicitly depends on $\epsilon_F$,
whereas Eq.(\call{HoppingHam}) does not.
This Hamiltonian reproduces the properties of the infinite system extremely
well even on a relatively small lattice.

If OBC's are used on this system, edge effects produce slowly decaying
Friedel-type oscillations in local proprieties, such as the density. PBC's
work much better, but still, the typical energy level spacing decays only as
$1/L$. SBC's concentrate more states at $\epsilon_F$ than elsewhere. The
advantages of this are apparent in Fig.~2, where
we plot the average kinetic energy per site, $\bra K \ket$, as a
function of the chemical potential, $\mu=\epsilon_{\rm F}$.
The choice of PBC's shows the presence of discontinuous jumps, typical of a
finite size system.
On the other hand, the use of SBC's eliminates the
discontinuities in $\bra K \ket$ already on a system as small as $L=30$ sites,
and agrees extremely well with the infinite system results.
The Friedel-like edge effects are also absent (not shown).

We next consider incommensurate spin-density-wave order in the positive-$U$
1-D Hubbard Hamiltonian\refto{Hubbard} within a mean-field approximation.
The Hubbard Hamiltonian is
$$
H=-\sum_{i,\sigma}t_{i}(c^{\dagger}_{i,\sigma}c_{i+1,\sigma} +
c^{\dagger}_{i+1,\sigma}c_{i,\sigma}) + \sum_i U_i n_{i\uparrow}
n_{i\downarrow}
-\sum_{i,\sigma}\mu_i n_{i\sigma}, \eqno(HubHam)
$$
which consists of a system of electrons
with an on-site interaction with coupling constant $U_i$. Here $t_{i}$
is the
nearest-neighbor hopping parameter between sites $i$ and $i+1$,
and $\mu_i$ is the chemical potential.
The $c^\dagger_{i\sigma}$ are fermion creation operators at site $i$ with spin
$\sigma$, and $n_{i\sigma}=c^\dagger_{i\sigma}c_{i\sigma}$.
Here $t_i/t$ is scaled according to the left-hand side of
Eq.(\call{fsmooth}) when
we use SBC's and $U_i/U=\mu_i/\mu={1\over{2t}}(t_{i-1}+t_i)$, where
$t$, $U$, and $\mu$ are the bulk values.

Applying the Hartree-Fock approximation, we
%approximation which neglects fluctuations of the $xy$ components of the spins.
rewrite the density operators as
$$
n_{i,\sigma}=\bra n_{i,\sigma} \ket +\delta n_{i,\sigma}=
\bra n_{i,\sigma} \ket + (n_{i,\sigma} - \bra n_{i,\sigma} \ket ).
\eqno(nfluct)
$$
We then insert Eq.(\call{nfluct}) in the Hamiltonian of Eq.(\call{HubHam}),
ignore terms quadratic in the density fluctuations, $\delta n_{i,\sigma}$,
and obtain the effective Hartree-Fock Hamiltonian,
$$
H_{\rm HF}=-\sum_{i,\sigma}t_{i}(c^{\dagger}_{i,\sigma}c_{i+1,\sigma} +
c^{\dagger}_{i+1,\sigma}c_{i,\sigma}) +
\sum_{i,\sigma} (U_i \bra n_{i,-\sigma}\ket -\mu_i ) n_{i,\sigma},
\eqno(HFHubHam)
$$
where we have dropped all constant terms.
This Hamiltonian can be easily diagonalized, and solutions can be
found self-consistently by iteration.
Previous studies using PBC's and OBC's have shown
that the Hamiltonian in Eq.(\call{HFHubHam}) has both spin and
charge incommensurate density waves.\refto{Schulz,IncommSpin}
Here we will show that the incommensurate
wavelength for the bulk can be already determined to high accuracy on a
small lattice using SBC's but not with standard boundary conditions.

In Fig.~3 we show the incommensurate spin-density-wave vector $q$ as a
function of the chemical potential, $\mu=\epsilon_{\rm F}$,
on a lattice with $L=30$ sites.
%Similarly to what we found for $\bra n\ket$ in Fig.~3,
We find that, when applying
PBC's to the system, $q$ takes only commensurate,
discrete values.
On the other hand, when considering SBC's
with all energy scales ($t_i/t$ and $U_i/U$)
decreasing on the right-
and left-most $10$ sites according to the smooth function defined in
Eq.(\call{fsmooth}), we see that $q$ increases smoothly with
$\mu$ in agreement with the infinite
lattice results, which were
derived from solving the system on larger lattices ($L=120,180$ sites)
with OBC's and SBC's and finding no changes in the results
upon increasing $L$ or changing types of boundary conditions. It is clear from
Fig.~3 that, even on a small lattice ($L=30$), SBC's
give results that are in good agreement with the results in the bulk.

To show that the application of SBC's is not only effective for
non-interacting systems or within mean-field theories, we studied the
Heisenberg chain using the density-matrix RG
approach.\refto{RGWhite,RGWhite1}
Here, we consider an antiferromagnetic $S=1/2$ Heisenberg chain described by
the
Hamiltonian,
$$
H=\sum_{i=1}^L J_i \vec{S_i} \cdot \vec{S_{i+1}}. \eqno(Heis)
$$
The Bethe ansatz exact solution to the model for the infinite system predicts
a ground state energy, $E_{0}=1/4-{\rm ln}~2$,
with $S_z^T=0$, where $S_z^T$ is the $z$ component of the total
spin.\refto{Bethe}
As expected from the valence bond picture
of the $S={1\over 2}$ chain,\refto{RGWhite1} the
density-matrix RG calculations show
that the effect of OBC's
causes a strong alternation in the bond-strength,
$\bra \vec{S}_i \cdot \vec{S}_{i+1} \ket$,
as a function of the site index, $i$, This alternation decays very slowly as
the size of the system is increased.\refto{RGWhite1}

We apply SBC's in order to eliminate, even on a
relatively small chain,
the bond-strength alternation
which is absent in the
infinite system.
For this purpose, we
choose $J_i/J$ according to the function on the left-hand side of
Eq.(\call{fsmooth}).
In Fig.~4 we show the bond energy per site,
$\bra \vec{S}_i \cdot \vec{S}_{i+1} \ket$
as a function of the site index, $i$, with OBC's and SBC's
on a chain with $L=60$ sites. It is clear that the strong bond alternation
present in the system with OBC's is strongly
suppressed when we apply SBC's.

In summary,
we have studied the effect of SBC's on one-dimensional
systems of interacting particles.
For all systems under consideration (non-interacting Fermi gas,
Hubbard model, and Heisenberg chain) and within all numerical techniques
used (exact diagonalization, mean-field self-consistent calculations,
numerical renormalization group) the use of SBC's
allows one to greatly reduce finite size effects (such as spatial fluctuations
and frustration) and extrapolate to the thermodynamic limit on relatively
small systems.
The use of SBC's can easily be extended to
quantum Monte Carlo techniques, and to systems of higher dimensionality, where
work is still under progress.

We would like to thank R.M.~Noack
for very helpful discussions. This work was supported by
the Office of Naval Research, grant
No. N00014-91-J-1143.
%The numerical calculations were performed primarily
%on the Cray Y-MP at the San Diego Supercomputer Center.
This work was also supported in part by the University of
California through an allocation of computer time.
\vfill\eject

\references

\refis{Bethe} H.A.~Bethe, {\it Z. Phys.} {\bf 71}, 205, (1931).\par
%\refis{Haldane} F.D.M.~Haldane, {\it Phys. Lett.} {\bf 93A}, 464, (1983).\par
\refis{Hardy}G.H.~Hardy, {\it Divergent Series} (Oxford, 1949).\par
\refis{selfdetbc}W.M.~Saslow, M.Gabay, and W.-M.~Zhang, \prl 68, 3627,
1992.\par
\refis{nebulabc}J.~Kolafa, {\it Molec. Phys.} {\bf 74}, 143, (1991).\par
\refis{Hubbard}J.~Hubbard, {\it Proc. Roy. Soc.} {\bf A276}, 238, (1963).\par
\refis{Schulz}H.J.~Schulz, \prl 64, 1445, 1990.\par
\refis{IncommSpin}M.~Kato, K.~Machida, H.~Nakanishi, and M.~Fujita, {\it J.
Phys. Soc. Jpn.} {\bf 59}, 1047, (1990).\par
\refis{RGWhite} S.R.~White, \prl 69, 2863, 1992.\par
\refis{RGWhite1} S.R.~White, to appear in {\it Phys. Rev. B}.\par

\endreferences

\figurecaptions
FIG.1 The smoothing function, $c_m$, as a function of the lattice
site, $m$. The solid line corresponds to Eq.(\call{ydef}) in the text, and the
squares correspond to the smoothing function derived from the
Borel transform defined through Eq.(\call{borelc}).
%Here we used $M=40$ with
%$M'=20$.

FIG.2 The average kinetic energy, $\bra K \ket$ as a function of the chemical
potential, $\mu$,
for the non-interacting one-dimensional tight-binding
chain with $L=30$ sites.
For SBC's the
smoothing occurs on the left- and right-most 10 sites.

FIG.3 The incommensurate spin-density-wave vector, $q$,
on a Hubbard chain as a function of the
chemical potential, $\mu$.
The chain has $L=30$ sites, the on-site repulsion is $U/t=2.0$,
and for the SBC's the
smoothing occurs on the left- and right-most 10 sites.
Here $q$ is rescaled by $L/\pi$
in order to show that with PBC's the spin-density-wave is
commensurate with the lattice.

FIG.4 The bond strength,
$\bra \vec{S}_i \cdot \vec{S}_{i+1} \ket$, as a function of the site
index, $i$, for an $S=1/2$ Heisenberg chain with $L=60$ sites and $m=128$
states kept.
For SBC's the smoothing occurs on the left- and right-most $10$ sites.
%The results are obtained using the numerical renormalization
%finite system method.
%The solid line is with OBC's, the open circles are with SBC's, and the
%horizontal dashed line is for the infinite system.

\endfigurecaptions

\end